\documentstyle[11pt,newpasp,twoside,epsf]{article}
\begin{document}

\title{Eclipsing binaries in open clusters. \\ I. V615\,Per and V618\,Per in h\,Persei}
\author{J.~Southworth, P.~F.~L.~Maxted and B.~Smalley}
\affil{Astrophysics Group, Keele University, Staffordshire, ST5~5BG, UK}

\begin{abstract}
Accurate physical parameters have been determined for two early-type detached eclipsing binaries in the open cluster h\,Persei (NGC 869). Masses accurate to 1.5\% are derived from high-resolution spectroscopy and radii accurate to 4--6\% have been obtained from fitting the existing light curves. The four stars are placed in the mass--radius plane and compared to the theoretical stellar models of the Granada Group. The best-fitting models have a low metallicity of $Z \approx 0.01$ and a high helium abundance of $Y = 0.34$. This is the first determination of the bulk metallicity of the Perseus Double Cluster. Recent studies have assumed a solar metallicity so their results should be reviewed. \end{abstract}

\section{Eclipsing binaries in open clusters}         \label{intro}

Theoretical stellar models are increasingly sophisticated and accurate. Different input physics and different sets of models often produce very similar predictions for the physical parameters of stars. This means that observational tests of stellar model physics require high quality data to determine the success of a particular model compared to other models. This is exacerbated by the number of physical properties which are not observable, allowing several quantities to be adjusted freely to fit the observed data, for example interior helium and metal abundances and the effect of different theories and treatments of convection. 

Detached eclipsing binaries (dEBs) can can be analysed to provide accurate and absolute values for the masses and radii of two stars of the same age and chemical composition (Andersen 1991), but the age and composition are not constrained by observations so can be varied freely when attempting to fit models to data. Photometric studies of open clusters are also commonly used to test theoretical predictions. Analyses of colour-magnitude diagrams can test quite subtle physics, but this method is usually limited in its usefulness due to degeneracies between the distance to the cluster, its age, metallicity and reddening effects, and the presence of unresolved multiple stars and field stars.

Eclipsing binaries in open clusters provide a way of combining these two methods to increase the number of physical properties which are directly observed. Accurate masses and radii of two stars are augmented by some knowledge of their age and chemical composition due to their membership of the cluster. A further advantage of having accurate radii is that the distance to each component of the dEB can be found via empirical calibrations of surface brightness (e.g., Barnes \& Evans 1976). This allows the determination of the distance to a cluster entirely independently of main sequence fitting or theoretical models.

The dEBs V615\,Persei and V618\,Persei were discovered by Krzesi\'nski, Pigulski \& Ko\l aczkowski (1999) in the well-studied open cluster h\,Persei. Their spectral types are B7\,V and A2\,V and their periods are 13.7 and 6.4 days respectively. They have proper motions (Uribe et~al.\ 2002), systemic velocities and photometric characteristics consistent with membership of the cluster. 

The Perseus Double Cluster is composed of the open clusters h\,Per and $\chi$\,Per, and contains a large number of unevolved early-type stars. The question of whether the clusters are physically related has been studied frequently throughout the last century. The results of recent photometric studies are converging to identical values of age ($\log\tau = 7.0 \pm 0.1$ years), and distance modulus ($11.70 \pm 0.05$\,mag), but final agreement has not yet been reached on whether the clusters are co-evolutionary (see Southworth, Maxted \& Smalley 2004).

\section{Observations and analysis}                   \label{obsanalysis}

\begin{figure} \plottwo{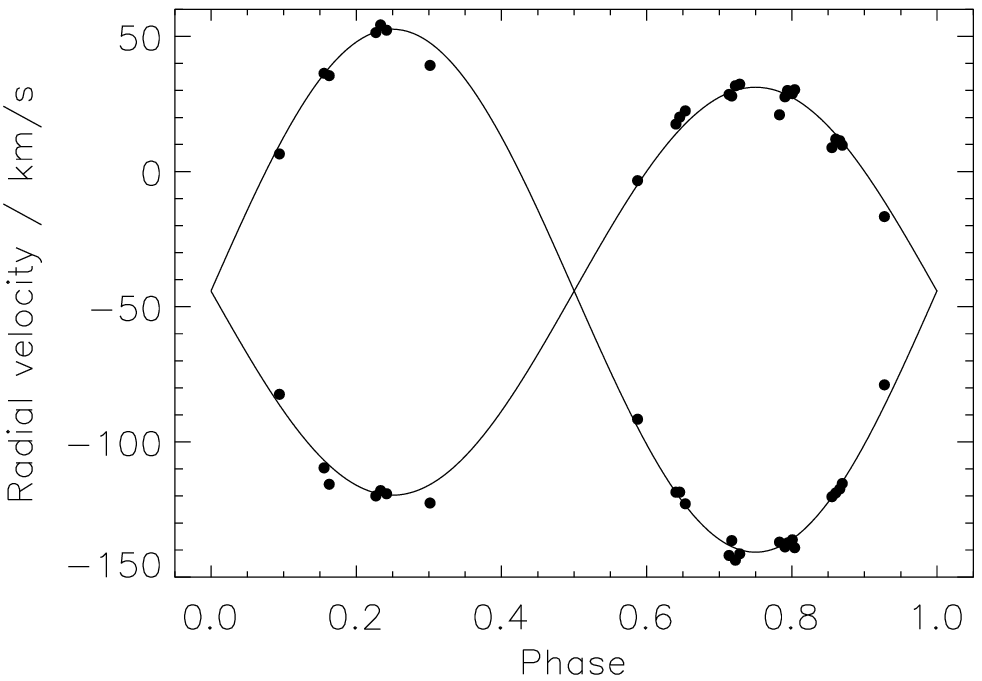}{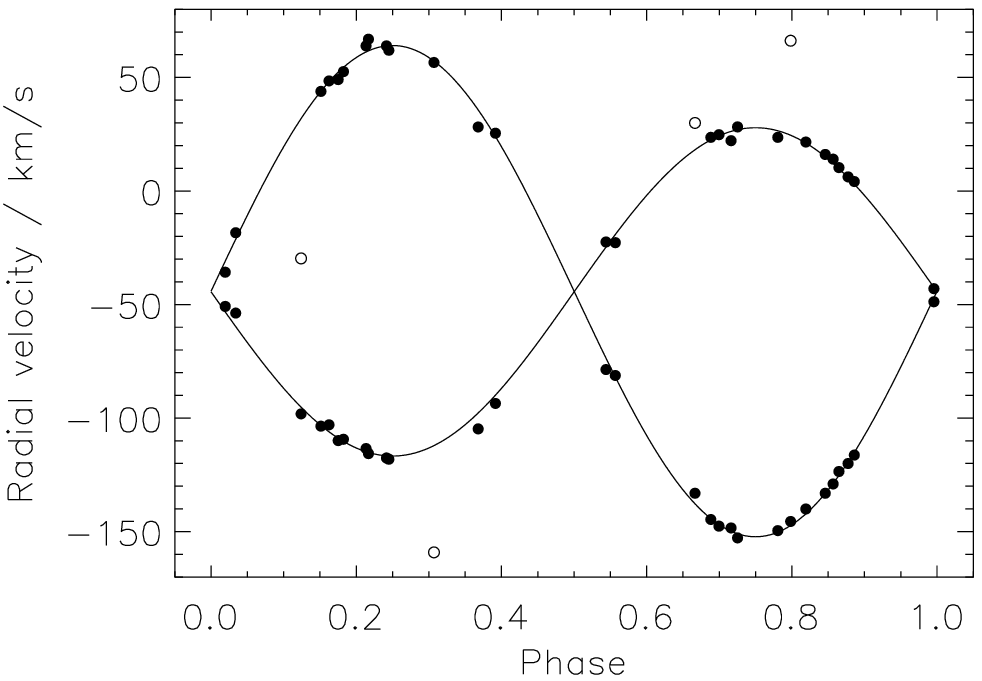} \caption{\label{orbits}Spectroscopic orbits of V615\,Per (left) and V618\,Per (right) fitted using {\sc sbop}. Open circles indicate rejected radial velocities.} \end{figure}

\begin{figure} \plottwo{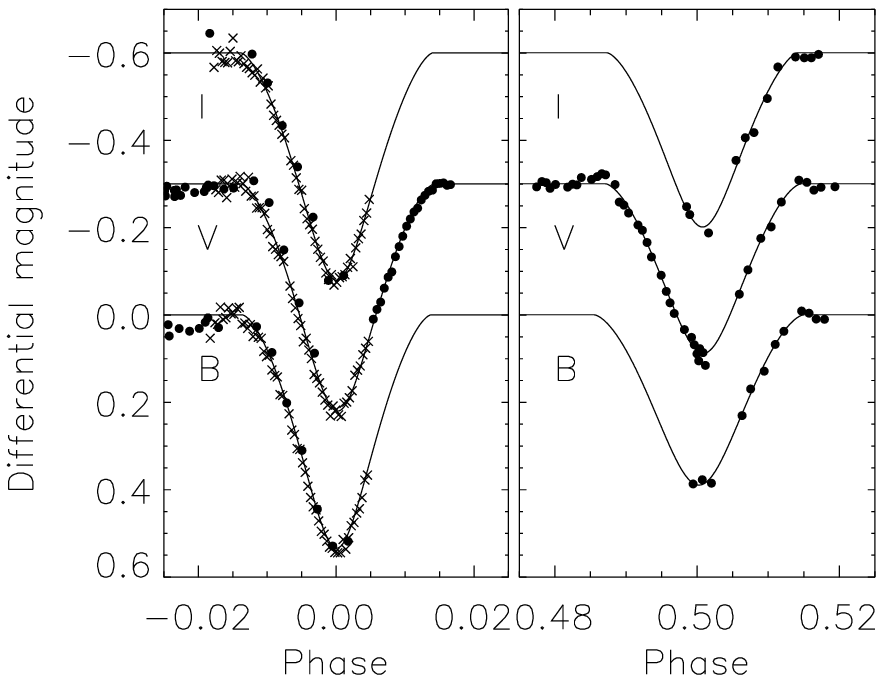}{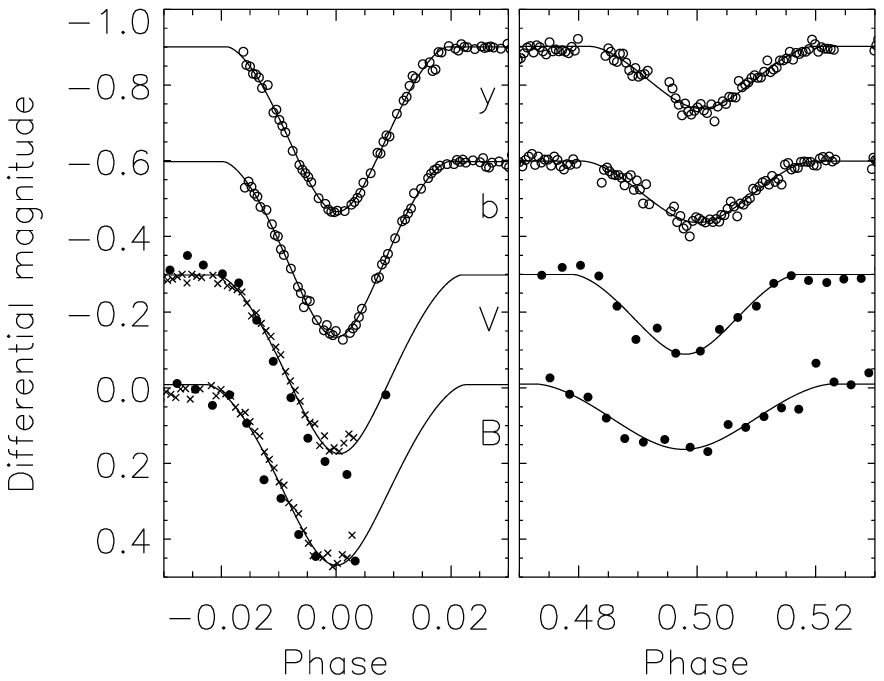} \caption{\label{lcs}Combined light curves of V615\,Per (left) and V618\,Per (right). The solid lines are the best fits calculated using {\sc ebop}.} \end{figure}

Grating spectra with a resolution of 0.2\,\AA\ were obtained with the Isaac Newton Telescope on La Palma. The spectral window included Mg\,{\sc ii} 4481\,\AA, known to give good radial velocities for early-type stars, and He\,{\sc i} 4471\,\AA, useful for the determination of effective temperatures. Exposure times of 1800\,s gave an approximate signal to noise per pixel of 50 for V615\,Per and 15 for V618\,Per.

The observed and disentangled spectra were analysed with synthetic spectra calculated using {\sc uclsyn} (Smalley et~al.\ 2001) and Kurucz {\sc atlas9} model atmospheres. Effective temperatures, rotational velocities, and light ratios were derived by fitting observations using least squares. All stars are slow rotators, with only the most massive star here rotating faster than synchronous.

Radial velocities were derived using the two-dimensional cross-correlation technique {\sc todcor} (Zucker \& Mazeh 1994) and synthetic templates generated using {\sc uclsyn}. Circular orbits were fitted to the derived radial velocities using the {\sc sbop} code (Fig.~\ref{orbits}). Uncertainties due to synthetic template mismatch were found using templates with a large range of effective temperature, rotational velocity and microturbulence velocity, and added in quadrature to the formal errors of the parameters of the spectroscopic orbit calculated by {\sc sbop}. 

The discovery light curves (Krzesi\'nski et~al.\ 1999) are the result of approximately 200 hours of observations in the broad-band $UBVI$ filters. Additional photometric data were obtained with the Jakobus Kapteyn Telescope, on La Palma, in $BVI$ for V615\,Per and in the Str\"omgren $b$ and $y$ filters for V618\,Per.

\begin{figure} \plotone{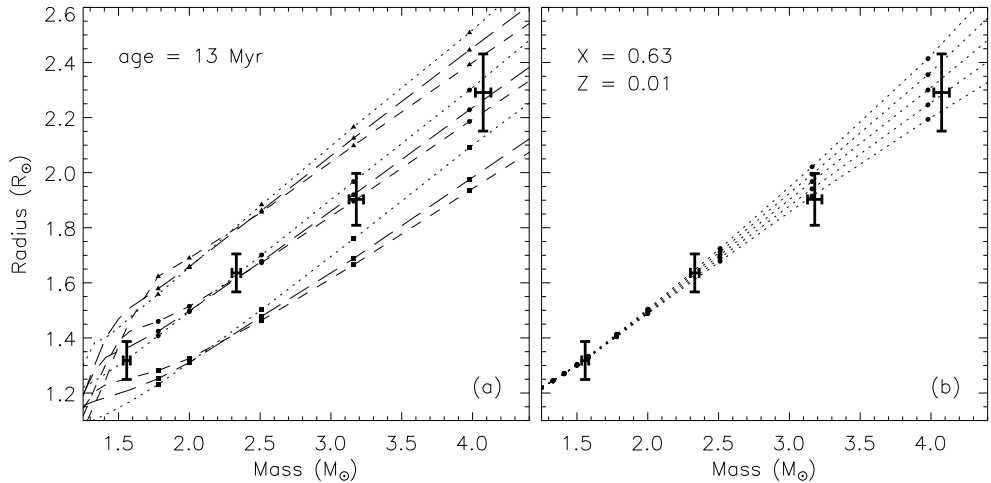} \caption{\label{modelfit}Comparison of the Granada models to the masses and radii of the components of V615\,Per and V618\,Per. (a) Models for $Z = 0.004$ (squares), $Z = 0.01$ (circles) and $Z = 0.02$ (triangles). Low, normal and high $X$ are plotted using short dashes, long dashes and dots. (b) ($X$,$Z$) = (0.63,0.01) model set plotted for ages of 3, 8, 13, 18, 23 Myr.} \end{figure}

The light curves were fitted using the {\sc ebop} code (Fig.~\ref{lcs}). A spectroscopic light ratio was used to break the degeneracy between the light ratio and the ratio of the radii for V615\,Per. The quoted uncertainties were derived from solutions of different-filter light curves and increased to include uncertainties in other quantities, e.g., orbital period and (theoretical) limb darkening coefficients. 

\section{Discussion and conclusions}                  \label{discussion}

We have derived absolute dimensions (Table~\ref{table}) for two early-type detached eclipsing binaries in the young open cluster h\,Persei (Southworth et~al.\ 2004). All four stars are slow rotators and both orbits have negligible eccentricity, despite the timescales for rotational synchronisation and orbital circularisation being much greater than the age of the cluster. This strongly supports the hypothesis of binary star formation by `delayed break-up' (Tohline 2002).

The stars have been plotted in the mass--radius plane in Fig.~\ref{modelfit} with the Granada theoretical models (Claret \& Gim\'enez 1998 and references therein) and a good fit is found for a chemical composition of ($X$,$Z$) = (0.63,0.01). Whilst several studies have derived atmospheric metal abundances of B\,stars in h and $\chi$ Per using high-resolution spectroscopy, this appears to be the first determination of the interior metallicity of stars in h\,Per using theoretical models. This metallicity is significantly below the solar metallicity assumed in recent photometric studies, suggesting that such studies could be systematically biased. 

Definitive light curves are needed for V615\,Per and V618\,Per to derive accurate (1\%) photometric parameters. Such data would allow a precise determination of the age, empirical distance, and metal and helium abundances, of h\,Per and a discriminate test of different stellar evolutionary models. 
%V621\,Per is a known dEB in the open cluster $\chi$\,Per and the determination of its absolute dimensions would allow us to plot six stars in the mass--radius diagram of h and $\chi$ Per. Initial analyses of spectra of V621\,Per suggest that they are single lined so we cannot determine absolute masses of the two components. The secondary star may be detectable using infrared spectroscopy and such observations could provide a very strong test of theoretical stellar evolutionary models.

\begin{table}
\caption{\label{table}Physical parameters of V615\,Persei and V618\,Persei.}
\begin{tabular}{l r@{\,$\pm$\,}l r@{\,$\pm$\,}l r@{\,$\pm$\,}l r@{\,$\pm$\,}l}
\tableline \tableline               & \multicolumn{2}{c}{V615\,Per A}   & \multicolumn{2}{c}{V615\,Per B}   & 
                                      \multicolumn{2}{c}{V615\,Per A}   & \multicolumn{2}{c}{V618\,Per B}         \\ 
\tableline
Mass ratio                        &\multicolumn{4}{c}{0.780 $\pm$ 0.010}&\multicolumn{4}{c}{0.668 $\pm$ 0.009}\\
$\frac{L_2}{L_1}$ (4481\,\AA)       & \multicolumn{4}{c}{0.47 $\pm$ 0.07} & \multicolumn{4}{c}{0.43 $\pm$ 0.18}   \\
%\tableline
Mass (M$_{\sun}$)                   & 4.08 & 0.06 & 3.18 & 0.05 & 2.33 & 0.03 & 1.56 & 0.03                       \\
Radius (R$_{\sun}$)                 & 2.29 & 0.14 & 1.90 & 0.09 & 1.64 & 0.07 & 1.32 & 0.07                       \\
$\log g$ (cgs)                      & 4.33 & 0.06 & 4.38 & 0.05 & 4.38 & 0.04 & 4.39 & 0.05                       \\
$T_{\rm eff}$ (K)                   & 15000 & 500 & 11000 & 500 & 11000 & 1000 & 8000 & 1000                      \\
$M_V$                               & 0.03 & 0.11 & 0.66 & 0.15 & 1.42 & 0.09 & 2.80 & 0.15                       \\
%\tableline
$V_{\rm rot}$ (km\,s$^{-1}$)        & 28 & 5 & 8 & 5 & 10 & 5 & 10 & 5                                            \\
$V_{\rm sys}$ (km\,s$^{-1}$)        & $-$44.3 & 0.7 & $-$44.1 & 0.5 & $-$44.4 & 0.8 & $-$44.3 & 0.5               \\
\tableline \tableline \end{tabular} \end{table}

\end{document}